\pdfoutput=1
\documentclass[aps,prl,reprint,superscriptaddress, showpacs]{revtex4-1}

\usepackage{graphicx}
\usepackage{natbib}
\usepackage[english]{babel}
\usepackage{amsmath}

\begin{document}


\title{Master-Slave Locking of Optomechanical Oscillators Over A Long Distance}


\author{Shreyas Y. Shah}
\author{Mian Zhang}
\affiliation{School of Electrical and Computer Engineering, Cornell University, Ithaca, New York 14853, USA}
\author{Richard Rand}
\affiliation{Department of Mathematics, Cornell University, Ithaca, New York 14853, USA}
\affiliation{Sibley School of Mechanical and Aerospace Engineering, Cornell University, Ithaca, New York 14853, USA}
\author{Michal Lipson}
\affiliation{School of Electrical and Computer Engineering, Cornell University, Ithaca, New York 14853, USA}
\affiliation{Kavli Institute at Cornell for Nanoscale Science, Ithaca, New York 14853, USA}



\begin{abstract}
Frequency-locking and other phenomena emerging from nonlinear interactions between mechanical oscillators are of scientific and technological importance. However, existing schemes to observe such behaviour are not scalable over distance. We demonstrate a scheme to couple two independent mechanical oscillators, separated in frequency by 80kHz and situated far from each other (3.2km), via light. Using light as the coupling medium enables this scheme to have low loss and be extended over long distances. This scheme is reversible and can be generalised for arbitrary network configurations.
\end{abstract}


\maketitle

\section{}
Frequency-locking between micromechanical oscillators is critical for RF communication and signal-processing applications \cite{strogatz2003, bregni2002, stephan1986}; however its scalability is limited by the fact that, in general, the oscillators are obliged to be in physical proximity in order to interact. Micromechanical oscillators can interact at the micron-scale via electronic coupling \cite{matheny2014} or a physical connection \cite{shim2007}. However, these schemes are fundamentally lossy over long distances, and therefore, are not scalable. Scaling up coupled mechanical oscillators to macro-scale networks \cite{tomadin2012, ludwig2013, heinrich2011} could potentially enable novel concepts in memory and computation \cite{mahboob2008, bagheri2011, hoppensteadt2001}, as well as provide a platform to put in practice many theories of nonlinear dynamics of coupled oscillators \cite{joshi2012, reddy1998}.

Interaction of mechanical oscillators through light could, in principle, help overcome this limitation, since light can propagate over long distances with minimal loss. Recent reports \cite{zhang2012, bagheri2013, shim2007} on frequency-locking between mechanical oscillators demonstrate interaction only over a few micrometers. In demonstrations of light-mediated coupling of two micromechanical oscillators \cite{zhang2012, bagheri2013}, both mechanical oscillators are coupled to the same optical cavity, limiting the kind of network topologies that can be used and how far the oscillators can be separated. 
\begin{center}
\begin{figure*}
\includegraphics[scale=0.185]{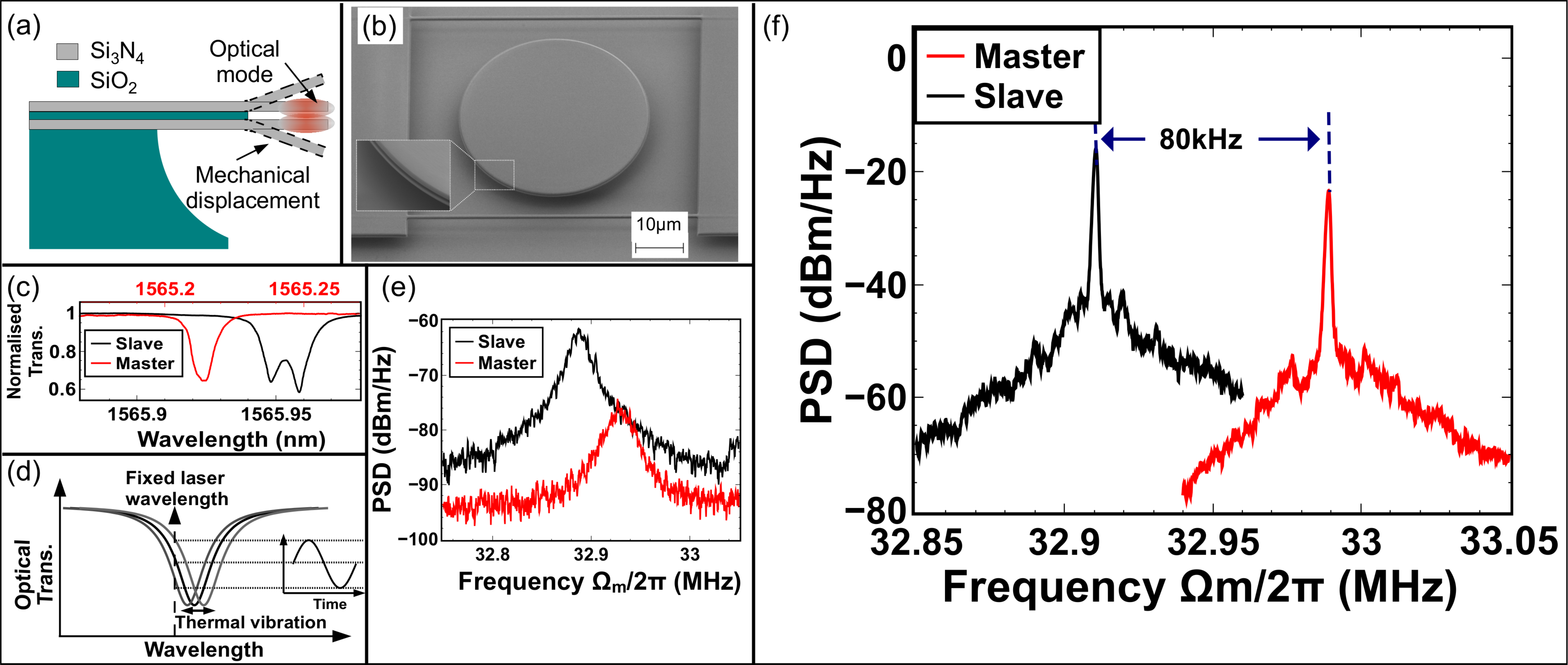}
\caption{\textbf{(a)} Schematic depiction of cross-section of the device, indicating the co-localisation of optical and mechanical resonances. The dotted line indicates the relative mechanical displacement between the two disks that influences the optical mode. \textbf{(b)} SEM image of the OM resonator. \textbf{(Inset)} Higher-magnification SEM image of the region highlighted, showing the double-microdisk structure. \textbf{(c)} Normalised transmission spectra of master and slave optical resonances. \textbf{(d)} Thermal vibration of the mechanical resonator causes the optical resonance to vibrate about a mean value, resulting in modulation of transmitted optical power \textbf{(e)} Power spectral density (PSD) of the modulation of the transmitted optical power due to thermally-induced mechanical vibration shows the natural frequency of the mechanical resonator. \textbf{(f)} PSD of master and slave oscillations. The oscillation peaks are offset by 80kHz.}
\end{figure*}
\end{center}

In this paper, we demonstrate a reconfigurable scheme to  couple, via light, two independent micromechanical oscillators  separated from each other by an effective path of 3.2km, in the master-slave configuration and show the ability to lock their oscillation frequencies. This coupling scheme is based on using light to send the information of the mechanical oscillations from the master oscillator to  the slave oscillator. It is facilitated by the fact that each oscillator is an an optomechanical oscillator (OMO), consisting of co-localised optical resonances and mechanical resonances that are coupled to each other (Eqs. \ref{eq1a}, \ref{eq1b}) \cite{vahala2007}. The mechanical resonator can be modelled as a damped simple harmonic oscillator with position `x', effective mass $m_{eff}$, frequency $\Omega_{m}$ and damping rate $\Gamma_{m}$. It is driven by its interaction with an optical force $F_{opt} = g_{om} \frac{\lvert a \rvert^{2}}{\omega}$, where $\lvert a \rvert^{2}$ is the energy in the optical cavity and $\omega$ is the laser frequency. $g_{om}$ indicates the strength of the interaction between optics and mechanics. The optical cavity can also be modelled as a damped oscillator, with a position-dependent frequency $(\omega_{0}+g_{om} x)$ and damping rate $\Gamma_{opt}$, and it is driven with a laser of power $\lvert s \rvert^{2}$, coupled to the cavity at the rate $\Gamma_{ex}$. The force on the mechanical resonator F$_{opt}$ can be controlled by changing the intracavity energy $\lvert a \rvert^{2} $,which is, in turn, affected by the laser power $\lvert s \rvert^{2}$. Any modulation of the laser power therefore couples to the mechanical resonator via the optical force F$_{opt}$ \cite{vahala2008}.
\begin{equation}\tag{1a}\label{eq1a}
\frac{da}{dt} = i((\omega - \omega_{0}) - g_{om}x)a - \Gamma_{opt}a + \sqrt{2\Gamma_{ex}}s
\end{equation}
\begin{equation}\tag{1b}\label{eq1b}
\frac{d^{2}x}{dt^{2}} + \Gamma_{m}\frac{dx}{dt}+\Omega_{m}^{2}x = \frac{F_{opt}[a]}{m_{eff}}
\end{equation}

The OMOs used for this demonstration each consist of two suspended Si$_{3}$N$_{4}$ microdisks stacked vertically (Figs. 1 (a), (b)). The optical and mechanical resonances under consideration are co-localised along the periphery of the structure. These structures are fabricated using e-beam lithography techniques \cite{zhang2012}. The top and bottom Si$_{3}$N$_{4}$ disks are nominally 250nm and 220nm thick and have a radius of 20$\mu$m. These disks are separated from each other by a 170nm thick SiO$_{2}$ sacrificial spacer layer. This stack rest	s on a 4$\mu$m thick SiO$_{2}$ support layer. These layers are partially etched away to release the periphery of these disks. This suspended structure supports optical whispering-gallery modes that are overlap with the edges of the top and bottom disks (Fig. 1(a)) \cite{zhang2012}. The optical resonance frequency of this structure is strongly dependent on the separation between the two disks. Relative motion (represented by Eq. 1(b)) between the two disks changes the resonance frequency at the rate of $g_{om}$=-2$\pi\cdot$49GHz/nm, as calculated from finite element simulations \cite{zhang2012}.

The two devices, when not coupled, oscillate at two distinct mechanical frequencies separated by 80kHz. In order to characterise the devices, light is coupled into each resonator with a tapered optical fiber. The transmission spectrum of the master OM resonator shows an optical resonance centered at $\sim$1565.22nm (Fig. 1(c)). Similarly, the slave OM resonator has an optical resonance centered at $\sim$1565.95nm (Fig. 1(c)). The splitting in the resonance is due to back-scattering induced lifting of degeneracy between the clockwise and counter-clockwise propagating modes \cite{vahala2002}. Thermal motion of the mechanical resonators modulates this transmission spectrum (Fig. 1(d)), which can be analysed with a spectrum analyser. The master is observed to have a mechanical resonance at 33.93MHz (Fig. 1(e)), with a linewidth of 16.39kHz, while the slave has a mechanical resonance centred at  32.82MHz (Fig. 1(e)), with a linewidth of 13.56kHz. When the optical resonances are excited with blue-detuned lasers ($\omega > \omega_{0}$), dynamical backaction \cite{vahala2007} amplifies mechanical motion. As input power is increased, this mechanical gain increases, until it overcomes intrinsic mechanical damping. At this point, each resonator becomes a self-sustaining oscillator \cite{vahala2007}. The master oscillates at 32.99MHz (Fig. 1(d)), and the slave oscillates independently at 32.91MHz (Fig. 1(d)), i.e. separated from the master by more than six times its natural  mechanical linewidth. Note that the oscillation frequencies for the oscillators are centred at a frequency slightly higher than that for the thermal motion of the respective resonator, due to the optical-spring effect \cite{vahala2007}.
\begin{figure}
\includegraphics[scale=0.5]{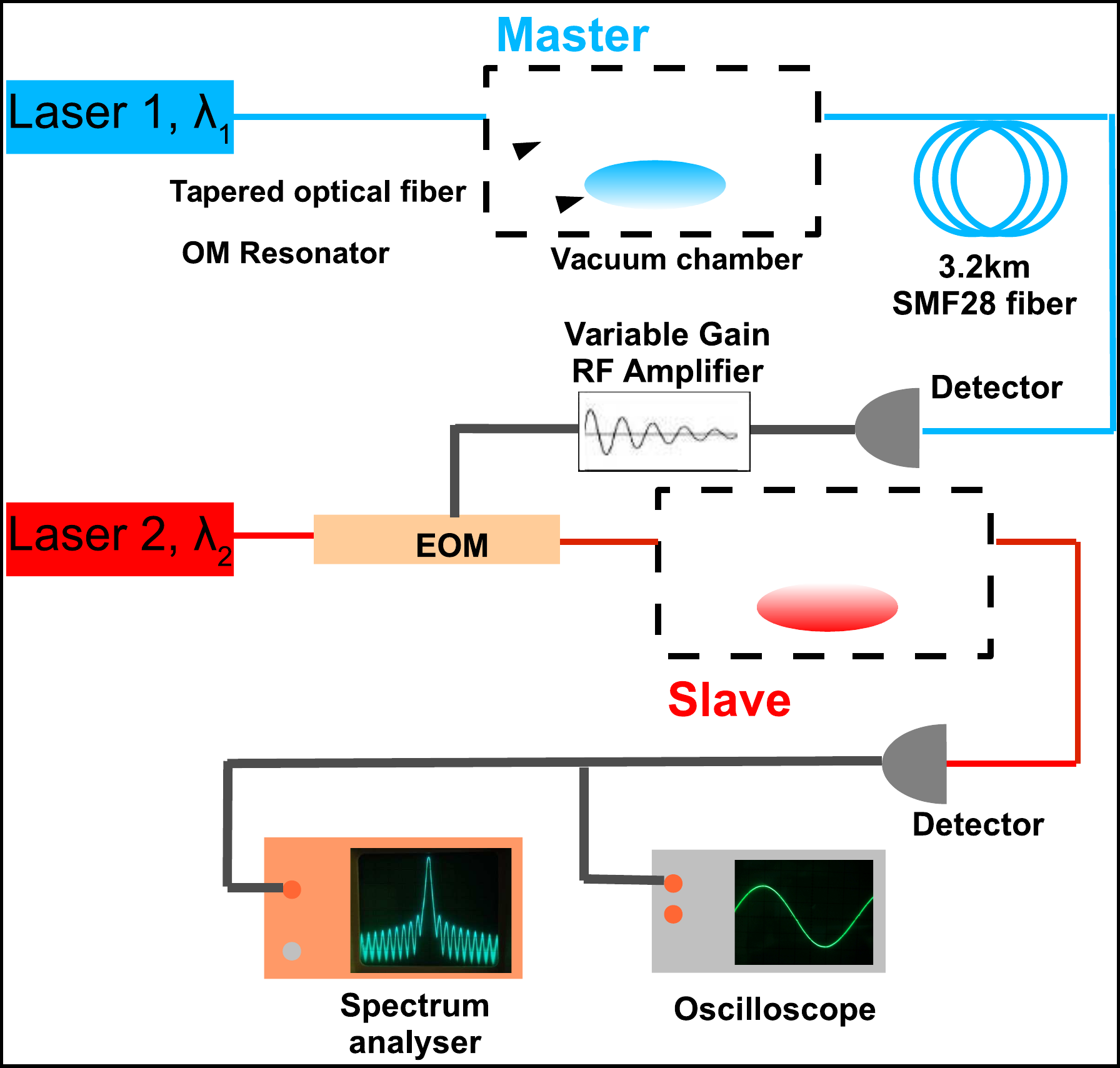}
\caption{Schematic of experimental setup to demonstrate master-slave locking. The two optomechanical (OM) resonators are driven by independent lasers. The optical signal from the master travels through 3.2km of fiber. The RF signal generated at the detector by the oscillations of the master modulate the laser driving the slave. The RF oscillations of the slave are analysed with a spectrum analyser and an oscilloscope.}
\end{figure}

To demonstrate long-distance locking, we couple the two OMOs in a master-slave configuration, via a 3.2km long optical fiber, with an electro-optic modulator that is driven by the master OMO and that modulates the laser driving the slave OMO (Fig. 2, Eq. 2). Each OMO is pumped by an independent laser. The signal transmitted from the master OMO carries information about its position `x$_{master}$'. It travels through a 3.2km long delay line before it is detected with a high-speed detector. The output of this detector carries the radio frequency (RF) oscillations, which are a function of the mechanical displacement x$_{master}$ of the master. The slave laser drive s$_{slave}$ is modulated by this signal from the master (Eq. 2). The output of the slave OMO is detected with another high-speed detector and analysed with a spectrum analyser and an oscilloscope.
\begin{equation}\tag{2}\label{eq2}
s_{slave} = s_{0,slave} + \gamma[f(x_{master})]
\end{equation}

The strength of coupling between the slave OMO and the output of the master OMO can be controlled by the modulation depth $\gamma$ of the electro-optic modulator driven by the master-oscillator. A voltage-controlled variable gain amplifier provides a gain between -26dB and +35dB to the RF oscillations coming from the detector of the master OMO, and thereby controls the modulation depth. This is reflected in the power spectral density (PSD) of oscillation peak of the master OMO (H$_{inj}$) as seen in the light transmitted from the slave OMO (Fig. 3(a)).
\begin{center}
\begin{figure*}
\includegraphics[scale=0.35]{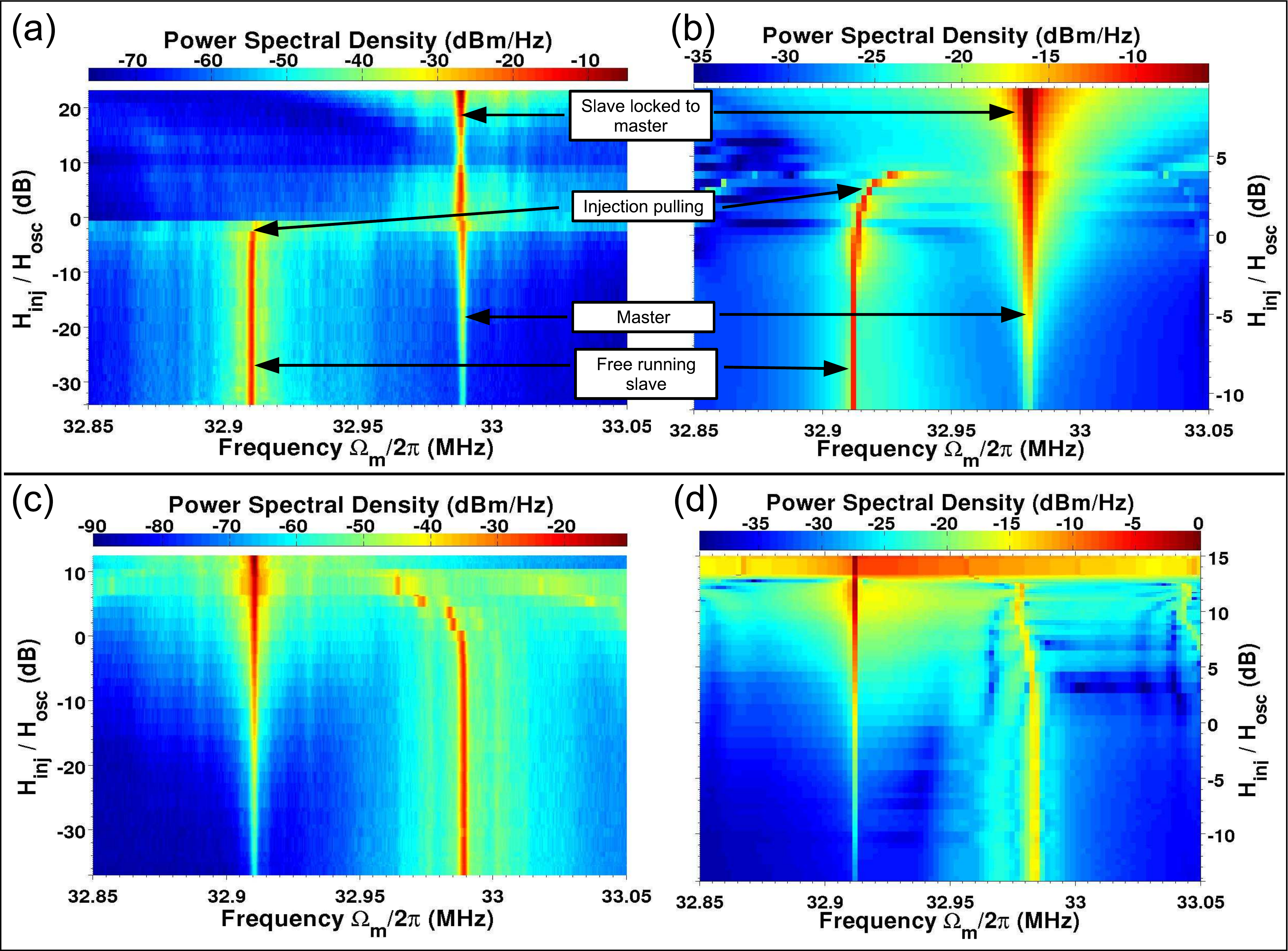}
\caption{\textbf{(a)} Spectrum of the power transmitted from the slave OMO for different injection ratios (H$_{inj}$/H$_{slave}$). \textbf{(b)} Numerical simulation of the power spectrum. \textbf{(c)}, \textbf{(d)} Same as \textbf{(a)} and \textbf{(b)}, respectively, only now measured by reversing the roles of master and slave.}
\end{figure*}
\end{center}

As we increase the coupling strength, we show that the slave OMO transitions from oscillating independently to being frequency-locked to the master OMO. The coupling strength, determined by the modulation depth, is measured in terms of the ratio of the power of injected oscillation signal (H$_{inj}$) to the power of the free-running slave oscillation (H$_{slave}$). When H$_{inj}$/H$_{slave}$ is small, the slave OMO oscillates at its own frequency, independently. The optical signal transmitted from the slave carries the slave oscillation peak, along with the modulation imparted on the laser (Fig. 3(a)). As the injection strength is increased, the slave oscillation frequency is pulled towards the master oscillation frequency. After a transition point (H$_{inj}$/H$_{slave}$ $\sim$ -2dB), the slave OMO spontaneously begins oscillating at the same frequency as the master OMO. 

We show that frequency locking can also occur when the roles of the slave and the master are reversed (Fig. 3(c)). As we increase the coupling strength, the new-slave spontaneously begins oscillating at the same frequency as the new-master after a transition point around H$_{inj}$/H$_{slave}$ $\sim$ 8dB. The difference in the locking strength for each of the oscillators can be attributed to the strongly nonlinear nature of these oscillators (\cite{zalalutdinov2003}, Supplemental Material \cite{supplementary2014}).
\begin{center}
\begin{figure*}
\includegraphics[scale=0.45]{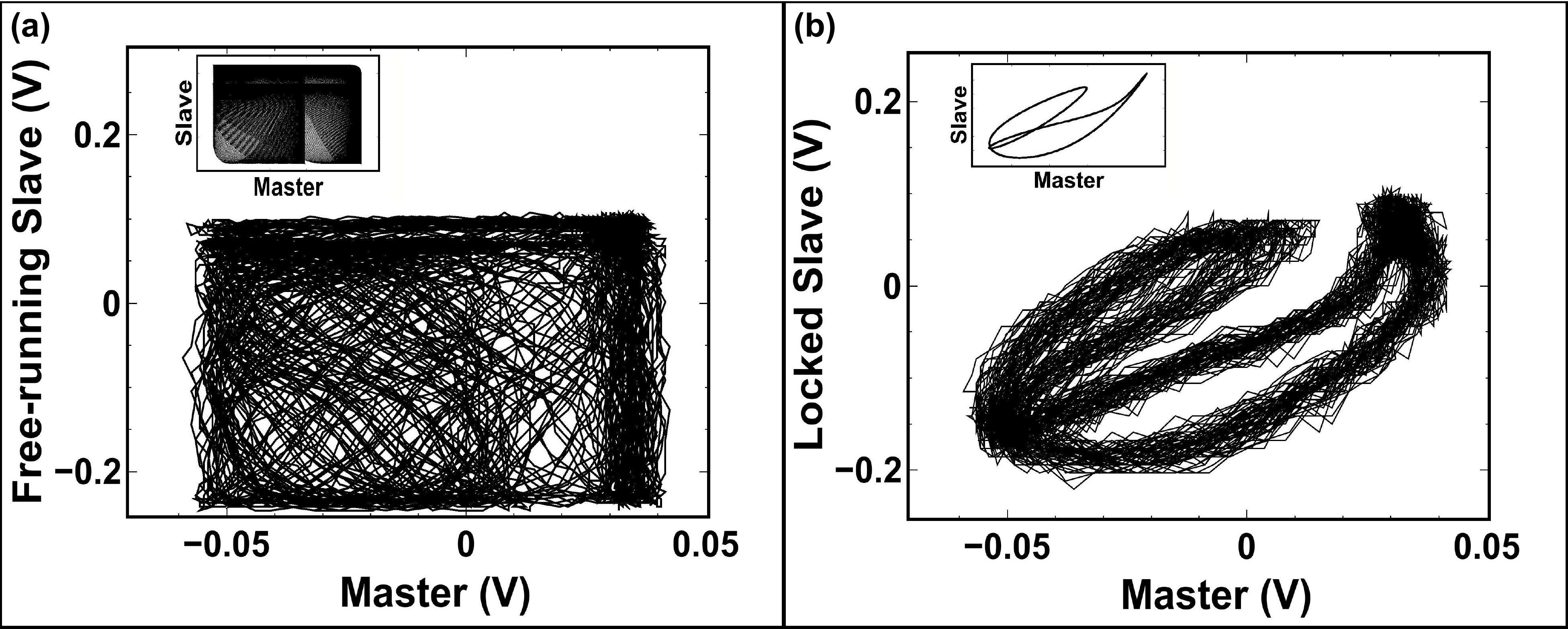}
\caption{Phase-portraits formed by the oscillation signals of the \textbf{(a)} free-running slave and \textbf{(b)} locked slave with the master oscillator, as measured with an oscilloscope, over more than 130 oscillation cycles.\textbf{(Insets)} Simulated phase-portraits}
\end{figure*}
\end{center}
We observe phase locking between the master and the slave oscillators when their frequencies lock. The locking-transition is associated with the establishment of a fixed phase-relationship between the master and the slave oscillations. We can observe the change in the phase-relationship upon locking between the master and slave oscillators by plotting the oscillation signal of the slave versus that of the master, over a duration long enough to accommodate phase drift. When the slave OMO is free-running, its phase is uncorrelated to the phase of the master OMO. As a result, for each point in the phase space of the master OMO, the phase of the slave OMO can take any value in its range (i.e. 0$^{0}$ – 360$^{0}$). This is reflected in the phase-portrait of the oscillations of the master OMO and slave OMO forming a filled-rectangle (Fig. 4(a)), over an extended period of time (4$\mu$s, i.e. more than 130 oscillation cycles) \cite{pikovsky2003}. When the slave OMO is locked to the master OMO, the phase difference between the two oscillations is fixed. This correlation between the phases of the two oscillators results in the X-Y trace of the oscillations (Fig. 4(b)) of the master and slave OMOs forming an open Lissajous figure \cite{pikovsky2003}.

Full numerical simulations of Eqs. 1 and 2 for the master and slave OMOs confirm the observation of locking (Fig. 3 (b), (d)). The dynamics of the slave OMO and the master OMO are simulated with experimentally-derived parameters. The set of coupled optical and mechanical equations (Eqs. 1) are numerically integrated using commercially available software [\cite{zhang2012}, Supplemental Material]. The optical-drive for the slave s$_{slave}$ (Eq. 2) in the simulation contains a signal $\gamma[f(x_{master})]$, which is proportional to the transmitted signal from the master OMO. As the gain is increased, the slave is locked to the oscillations of the master OMO. The simulations also reproduce, qualitatively, major features of the dynamics, including injection-pulling \cite{razavi2004}.

Our demonstration of master-slave locking of two OMOs separated by kilometers of fiber utilises a reconfigurable coupling scheme that can be easily extended to include mutual coupling between the two oscillators as well as to implementing a large network of oscillators with arbitrary network topologies. The ability to tune the coupling strength arbitrarily enables access to various regimes of nonlinear dynamics of such oscillator networks.

The authors gratefully acknowledge support from DARPA for award $\#$W911NF-11-1-0202 supervised by Dr. Jamil Abo-Shaeer. The authors also acknowledge Applied Optronics. This work was performed in part at the Cornell NanoScale Facility, a member of the National Nanotechnology Infrastructure Network, which is supported by the National Science Foundation (Grant ECCS-0335765). This work made use of the Cornell Center for Materials Research Facilities supported by the National Science Foundation under Award Number DMR-1120296. The authors also acknowledge Prof. Paul McEuen for use of lab facilities.
\bibliography{MasterSlaveLocking}

\end{document}